\documentclass[12pt,journal,a4paper,draftcls,oneside,onecolumn]{IEEEtran}
\usepackage{bbm}
\usepackage[compress]{cite}
\usepackage{stmaryrd}
\usepackage{mathrsfs}
\usepackage{amsfonts}
\usepackage{amsmath}
\usepackage{cases}
\usepackage{txfonts}
\usepackage{amssymb}
\usepackage{epsfig}
\newcommand{\fat}[1]{\mbox{\boldmath$#1$}}
\newtheorem{theorem}{Theorem}

\newtheorem{lemma}{Lemma}

\newtheorem{remark}{Remark}

\newcommand{\myQED}{\mbox{}\hfill{$\Box$}}

\ifCLASSINFOpdf
\else
\fi
\begin{document}
%
\title{Analysis of Finite Field Spreading \\for Multiple-Access Channel}
%
%
%

\author{Guanghui~Song,~\IEEEmembership{Student Member,~IEEE,}
Yuta Tsujii,~\IEEEmembership{Student Member,~IEEE,}
        Jun Cheng,~\IEEEmembership{Member,~IEEE,}
        and~Yoichiro~Watanabe,~\IEEEmembership{Member,~IEEE}}
\maketitle

\begin{abstract}
Finite field spreading scheme is proposed for a synchronous
multiple-access channel with Gaussian noise and equal-power users.
For each user, $s$ information bits are spread \emph{jointly} into a
length-$sL$ vector by $L$ multiplications on GF($2^s$). Thus, each
information bit is dispersed into $sL$ transmitted symbols, and the
finite field despreading (FF-DES) of each bit can take advantage of
$sL$ independent receiving observations. To show the performance
gain of joint spreading quantitatively, an extrinsic information
transfer (EXIT) function analysis of the FF-DES is given. It shows
that the asymptotic slope of this EXIT function increases as $s$
increases and is in fact the absolute slope of the bit error rate
(BER) curve at the low BER region. This means that by increasing the
length $s$ of information bits for joint spreading, a larger
absolute slope of the BER curve is achieved. For $s, L\geq2$, the
BER curve of the finite field spreading has a larger absolute slope
than that of the single-user transmission with BPSK modulation.
\end{abstract}


%
\IEEEpeerreviewmaketitle

\section{Introduction}

In a $K$-user multiple-access channel (MAC), each user regards the
signals of other users as interference. When the number of users $K$
is large, each user has a very low signal-to-interference-and-noise
ratio (SINR) \cite{verylowrate}. For this reason, spreading is
usually employed as an SINR amplifier for each user, such as the
conventional code-division multiple-access (CDMA) \cite{cdma} and
interleave-division multiple-access (IDMA) systems
\cite{IDMA,idmaj,analysisIDMA,spreaddesign}.

In the CDMA and IDMA systems, each information bit is spread
\emph{independently}. Specifically, in the CDMA system each
information bit is spread by multiplying a binary vector into a
length-$L$ bit-vector, which is then interleaved by a bit-level
interleaver. Since each information bit has $L$ independent
receiving observations, the despreading (DES) will output an
ameliorated signal with a sufficient large SINR for a further outer
decoding (if there is a channel code as an outer code). In the IDMA
system, a chip-level interleaving is jointly performed on multiple
bit-vectors instead of the bit-level interleaving in the CDMA system
\cite{idmaj}. As the number of bit-vectors for interleaving is
large, the IDMA in fact becomes a multi-user sparse-graph code that
is appropriate for iterative decoding \cite{spreaddesign}. For this
reason, the IDMA under an iterative chip-by-chip decoding provides a
lower bit error rate (BER) than the conventional CDMA
\cite{IDMA}\cite{spreaddesign}.
Simulations in \cite{IDMA} showed that at the low BER region, the
BER of the uncoded IDMA system can converge to that of single-user
transmission with BPSK modulation. The BER of single-user
transmission with BPSK modulation is the performance upper bound for
both of the CDMA and IDMA systems, since in both systems the
independent spreading of each information bit implies that each user
employs a repetition code.

In this paper, we propose a finite field spreading scheme for a
synchronous MAC with Gaussian noise and equal-power users. For each
user, every $s$ information bits are spread \emph{jointly} into a
length-$sL$ vector by $L$ multiplications on GF($2^s$). A chip-level
interleaving is then performed to generate the transmitted vector to
the MAC. At the receiver, a multi-user iterative decoding is
performed on a single factor graph to recover the information vector
of each user. In our scheme, due to the joint spreading, each
information bit is dispersed into $sL$ transmitted symbols, and the
finite field despreading (FF-DES) of each bit can take advantage of
$sL$ independent receiving observations. To show the performance
gain of joint spreading quantitatively, we analyze the extrinsic
information transfer (EXIT) function of the FF-DES. We show that the
asymptotic slope of this EXIT function increases as $s$ increases
and is in fact the absolute slope of the BER curve at the low BER
region. This means that by increasing the length $s$ of information
bits for joint spreading, a larger absolute slope of the BER curve
is achieved. For $s, L\geq2$, the BER curve of the finite field
spreading has a larger absolute slope than that of the single-user
transmission with BPSK modulation.

\begin{figure}
\includegraphics[width=3.5 in]{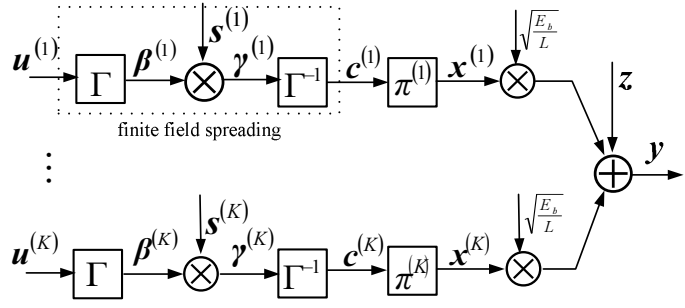}
\centering
%
\caption{Transmitter of $K$-user finite field spreading
multiple-access system.} \label{fig:transmitter}
\end{figure}
\section{Transmitter Model} \label{sec:tansmitter}
Figure~\ref{fig:transmitter} illustrates a block diagram of the
transmitter of $K$-user finite field spreading multiple-access
system. The length-$sN$ information vector
$\fat{u}^{(k)}=(u^{(k)}_1,u^{(k)}_2,...,u^{(k)}_{sN}),
u_i^{(k)}\in\{+1,-1\}\triangleq\mathcal {X}, 1\leq k\leq K$, of user
$k$ is first mapped into a length-$N$ vector
$\fat{\beta}^{(k)}=(\beta^{(k)}_1,\beta^{(k)}_2,...,\beta^{(k)}_{N})$
over GF($2^s$), i.e., every $s$ information bits are mapped into a
field element by mapping $\Gamma: \mathcal {X}^s\rightarrow
\textrm{GF}(2^s)$. Here GF$(2^s)=\{0,1,\alpha...,\alpha^{2^s-2}\}$
is a finite field with a primitive element $\alpha$, and $\Gamma$
can be an arbitrary bijection from $\mathcal {X}^s$ to GF($2^s$).
Each element $\beta^{(k)}_j, 1\leq j\leq N$, in $\fat{\beta}^{(k)}$
is spread by length-$L$ spreading vector $\fat{s}^{(k)}$ over
GF($2^s$) into
$\beta^{(k)}_j\fat{s}^{(k)}=(\beta^{(k)}_js^{(k)}_1,\beta^{(k)}_js^{(k)}_2,...,\beta^{(k)}_js^{(k)}_{L})$,
where the multiplication is on GF($2^s$). Here $\fat{s}^{(k)}$ can
be an arbitrary vector over GF($2^s$) with $s^{(k)}_\ell\neq 0,
\ell=1,2,...,L$. The output field vector after multiplication,
denoted as
$\fat{\gamma}^{(k)}=(\gamma^{(k)}_1,\gamma^{(k)}_2,...,\gamma^{(k)}_{NL})$,
is demapped into binary vector
$\fat{c}^{(k)}=(c^{(k)}_1,c^{(k)}_2,...,c^{(k)}_{sNL})$, i.e., each
field element is demapped into $s$ bits by demapping $\Gamma^{-1}:
\textrm{GF}(2^s)\rightarrow \mathcal{X}^s$, where $\Gamma^{-1}$ is
the inverse transform of $\Gamma$. Vector $\fat{c}^{(k)}$, referred
to as chip vector, is interleaved by a length-$sNL$ chip-level
interleaver $\pi^{(k)}$ and is multiplied by amplitude
$\sqrt{\frac{E_b}{L}}$ to generate the transmitted vector
$\sqrt{\frac{E_b}{L}}\fat{x}^{(k)}=(\sqrt{\frac{E_b}{L}}x^{(k)}_1,\sqrt{\frac{E_b}{L}}x^{(k)}_2,...,\sqrt{\frac{E_b}{L}}x^{(k)}_{sNL}),x^{(k)}_t\in\mathcal
{X}$, to the Gaussian MAC. Here $E_b$ is the energy per information
bit, and $\frac{E_b}{L}$ is the symbol energy per transmission due
to code rate $\frac{1}{L}$ of each user. The chip-level interleaving
should be different for each user as in the IDMA system
\cite{IDMA,idmaj,analysisIDMA,spreaddesign}.


The receiver receives a superimposed signal vector
$\fat{y}=(y_1,y_2,...,y_{sNL})$ with
\begin{equation} \label{eq:receive}
y_t=\sum_{k=1}^K\sqrt{\frac{E_b}{L}}{x}^{(k)}_t+z_t,\ t=1,2,...,sNL
\end{equation}
where $z_t$ is a zero-mean Gaussian variable with one-side power
spectral density $N_0$. Here, we assume that the symbols from the
$K$ users are synchronous. An iterative multi-user decoding is
performed to recover information vectors $\fat{u}^{(k)}, k=1,...,K$.

To simplify the description, the mapping, multiplication of
spreading vector, and demapping in the dotted box in
Fig.~\ref{fig:transmitter} is referred to as finite field spreading.
It should be noted that by the finite field spreading over
GF$(2^s)$, the spreading of $s$ information bits is performed
jointly, and each information bit is dispersed into $sL$ transmitted
symbols. Thus, the FF-DES of each information bit can take advantage
of $sL$ independent receiving observations. Conventional spreading
schemes, such as the CDMA and IDMA where the information bit is
spread independently, are special cases of the finite field
spreading with $s=1$.

\section{Iterative Decoding on Factor Graph}\label{sec:decoding}
Due to the chip-level interleaving, as length $N$ is large, the
$K$-user finite field spreading multiple-access can be regarded as a
$K$-user sparse-graph code which can be decoded iteratively on a
single factor graph. In this section, we give an iterative decoding
algorithm of the finite field spreading multiple-access on a single
factor graph. In Section~\ref{sec:factor}, we represent the $K$-user
finite field spreading and the MAC by a single factor graph. In
Section~\ref{sec:iterative}, we give the iterative decoding
algorithm.

\subsection{Factor Graph}\label{sec:factor}
\begin{figure}
\includegraphics[width=3.3 in]{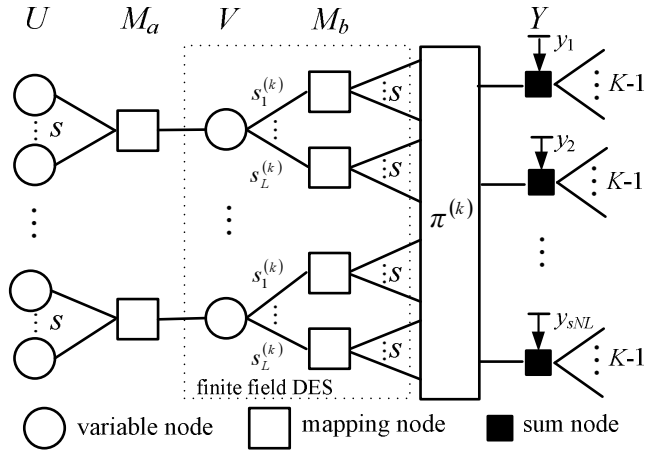}
\centering
%
\caption{Factor graph of the finite field spreading and the MAC for
user $k$.} \label{fig:factorfinite}\vspace{-3mm}
\end{figure}
Since each user has the same encoding process, the factor graph of
each user is the same. We only illustrate the factor graph of user
$k$ in Fig.~\ref{fig:factorfinite}. There are three kinds of nodes
in the graph: variable, mapping, and sum nodes. Let $U,\ M_a,\ V,\
M_b$, and $Y$ be five node sets that include the nodes below letters
``$U$," ``$M_a$," ``$V$," ``$M_b$," and ``$Y$" in
Fig.~\ref{fig:factorfinite}. The variable node in $U$ corresponds to
an information bit. The variable node in $V$ corresponds to a symbol
in GF($2^s$). The mapping node in $M_a$ or $M_b$ denotes a mapping
relation between a GF($2^s$) symbol and $s$ bits. The sum node in
$Y$, associated with a received symbol, denotes a superposition of
the transmitted symbols from $K$ users. Note that the sum node
connects the remaining $K-1$ user's factor graphs. Edges connecting
variable nodes in $V$ and mapping nodes in $M_b$ are labeled by
elements of spreading vector $\fat{s}^{(k)}$. During encoding and
decoding, the message should multiply the labeled element or its
inverse when
passing across one of these edges. 

\subsection{Iterative Decoding}\label{sec:iterative}
The iterative decoding is performed on the factor graph in
Fig.~\ref{fig:factorfinite} and is accomplished by efficient local
decoding at all the nodes and interactions. The input and output of
local decoding at each node is denoted by a log-likelihood ratio
(LLR). Since the decoding of each user is performed in parallel and
is the same, we only give decoding algorithm for user $k$.

A single decoding iteration includes the local decoding at the sum
node in $Y$, deinterleaving, and the FF-DES in
Fig.~\ref{fig:factorfinite}. For the local decoding at the sum node
in $Y$, we employ the low-complexity elementary signal estimation
(ESE) algorithm given in \cite{IDMA,idmaj,analysis IDMA}. For the
FF-DES, we give a maximum a posteriori probability (MAP) decoding
algorithm. After a number of iterations, a hard decision is
performed to recover the information vector.



\subsubsection{Sum Node Decoding (ESE)}
Let $L^{a}(x^{(i)}_t), 1\leq i\leq K,i\neq k$, denote a priori LLR
of $x^{(i)}_t$ in (\ref{eq:receive}) to the $t$-th, $1\leq t\leq
sNL$, sum node in $Y$. By regarding $\sum_{i=1,i\neq
k}^K\sqrt{\frac{E_b}{L}}x^{(i)}_t$ as a Gaussian variable, this sum
node performs a local decoding and outputs an extrinsic LLR of
$x^{(k)}_t$ as \cite{IDMA,idmaj,analysis IDMA}
\begin{eqnarray}
\!\!\!\!\!\!\!\!\!\!\!\!\!\!\!\!\!\!\!\!L^e(x^{(k)}_t)&&\!\!\!\!\!\!\!\!\!\!=\log\frac{\textrm{Pr}(x^{(k)}_t=+1|y_t)}{\textrm{Pr}(x^{(k)}_t=-1|y_t)}\nonumber\\
\!\!\!\!\!\!\!\!\!\!\!\!\!\!\!\!\!\!\!\!&&\!\!\!\!\!\!\!\!\!\!=\frac{2\sqrt{\frac{E_b}{L}}\left(y_t-\sqrt{\frac{E_b}{L}}\sum_{i=1,i\neq
k}^K\tanh(\frac{L^{a}(x^{(i)}_t)}{2})\right)}{\frac{E_b}{L}\sum_{i=1,i\neq
k}^K\left(1-(\tanh(\frac{L^{a}(x^{(i)}_t)}{2}))^2\right)+\frac{N_0}{2}}\label{eq:ESE}
\end{eqnarray}
which will be deinterleaved as a priori LLR for the FF-DES.

\subsubsection{FF-DES}\label{sec:DES}
After deinterleaving, we obtain a priori LLR $L^a(c^{(k)}_t), 1\leq
t\leq sNL$, for each chip of user $k$. The FF-DES will calculate an
extrinsic LLR $L^e(c^{(k)}_t)$ based on the priori LLRs
$L^a(c^{(k)}_t), 1\leq t\leq sNL$, by the local decoding at the
nodes in $M_b\rightarrow V\rightarrow M_b$, subsequently.

Since the decoding operation for each chip is the same, to simplify
our description, we only introduce the calculation of extrinsic LLR
$L^e(c^{(k)}_{(\ell-1)s+n}),1\leq\ell\leq L,1\leq n\leq s$,
performed on the subgraph associated with the first variable node in
$V$ of Fig.~\ref{fig:factorfinite}.

Let $(L^a(c^{(k)}_{(i-1)s+1}),...,L^a(c^{(k)}_{is}))$ denote the $s$
chip-LLR inputs to the $i$-th, $1\leq i\leq L$, mapping node in
$M_b$ and $\fat{y}^\prime_i=(y^\prime_{(i-1)s+1},...,y^\prime_{is})$
denote the associated received symbols
($\fat{y}^\prime=(y^\prime_1,...,y^\prime_{sNL})$ is the
deinterleaved form of $\fat{y}$). This mapping node performs a local
decoding to transform chip LLRs to field symbol LLR vector
$\fat{L}^a(\gamma^{(k)}_i)=(L^a_{0}(\gamma^{(k)}_i),L^a_{1}(\gamma^{(k)}_i),...,L^a_{\alpha^{2^s-2}}(\gamma^{(k)}_i))$
of $\gamma^{(k)}_i$ as
\begin{eqnarray}
L^a_\lambda(\gamma^{(k)}_i)&&\!\!\!\!\!\!\!\!=\log\frac{\textrm{Pr}(\gamma^{(k)}_i=\lambda|\fat{y}^\prime_i)}{\textrm{Pr}(\gamma^{(k)}_i=0|\fat{y}^\prime_i)}\nonumber\\
&&\!\!\!\!\!\!\!\!=\log\frac{\prod_{m=1}^s{\textrm{Pr}(c^{(k)}_{(i-1)s+m}=\Gamma^{-1}_m(\lambda)|y^\prime_{(i-1)s+m})}}{\prod_{m=1}^s{\textrm{Pr}(c^{(k)}_{(i-1)s+m}=\Gamma^{-1}_m(0)|y^\prime_{(i-1)s+m})}}\nonumber\\
&&\!\!\!\!\!\!\!\!=\log\frac{\prod_{m=1}^s\frac{\textrm{Pr}(c^{(k)}_{(i-1)s+m}=\Gamma^{-1}_m(\lambda)|y^\prime_{(i-1)s+m})}{\textrm{Pr}(c^{(k)}_{(i-1)s+m}=-1|y^\prime_{(i-1)s+m})}}{\prod_{m=1}^s\frac{\textrm{Pr}(c^{(k)}_{(i-1)s+m}=\Gamma^{-1}_m(0)|y^\prime_{(i-1)s+m})}{\textrm{Pr}(c^{(k)}_{(i-1)s+m}=-1|y^\prime_{(i-1)s+m})}}\nonumber\\
&&\!\!\!\!\!\!\!\!=
\log\frac{\prod_{m=1}^s\left(\frac{\textrm{Pr}(c^{(k)}_{(i-1)s+m}=+1|y^\prime_{(i-1)s+m})}{\textrm{Pr}(c^{(k)}_{(i-1)s+m}=-1|y^\prime_{(i-1)s+m})}\right)^{\frac{(1+\Gamma^{-1}_m(\lambda))}{2}}}{\prod_{m=1}^s\left(\frac{\textrm{Pr}(c^{(k)}_{(i-1)s+m}=+1|y^\prime_{(i-1)s+m})}{\textrm{Pr}(c^{(k)}_{(i-1)s+m}=-1|y^\prime_{(i-1)s+m})}\right)^{\frac{(1+\Gamma^{-1}_m(0))}{2}}}\nonumber\\
&&\!\!\!\!\!\!\!\!=\sum_{m=1}^s\!\frac{\Gamma^{-1}_m(\lambda)\!-\!\Gamma^{-1}_m(0)}{2}L^a(c^{(k)}_{(i-1)s+m}),\
\lambda=0,1,...,\alpha^{2^s-2} \label{eq:bit-symbol}
\end{eqnarray}
where $\Gamma^{-1}_m(\lambda)$ takes the $m$-th bit of the demapped
vector $\Gamma^{-1}\!(\lambda)$. Here we use a posteriori
probability of zero element 0 as the denominator in a field symbol
LLR.

Let $\fat{y}_{\nmid
\ell}^\prime=(\fat{y}^\prime_{1},...,\fat{y}^\prime_{l-1},\fat{y}^\prime_{l+1},...,\fat{y}^\prime_{L})$
denote the joint vector of $\fat{y}^\prime_{i},i=1,...,L,i\neq
\ell$. Similarly, we have denotation $\fat{s}^{(k)}_{\nmid
\ell}=(s^{(k)}_1,...,s^{(k)}_{\ell-1},s^{(k)}_{\ell+1},...,s^{(k)}_L)$.
Based on the $L-1$ field symbol LLR vectors
$\fat{L}^a(\gamma^{(k)}_{i}),1\leq i\leq L,i\neq\ell$, the first
variable node in $V$ performs a local decoding to calculate the
extrinsic LLR vector
$\fat{L}^e(\gamma^{(k)}_{\ell})=(L^e_{0}(\gamma^{(k)}_{\ell}),L^e_{1}(\gamma^{(k)}_{\ell}),$
$...,L^e_{\alpha^{2^s-2}}(\gamma^{(k)}_{\ell}))$ of
$\gamma^{(k)}_{\ell}$ as
\begin{eqnarray}
\!\!\!\!\!\!\!\!L^e_\lambda(\gamma^{(k)}_{\ell})&&\!\!\!\!\!\!\!\!=\log\frac{\textrm{Pr}(\gamma^{(k)}_{\ell}=\lambda|\fat{y}_{\nmid \ell}^\prime)}{\textrm{Pr}(\gamma^{(k)}_{\ell}=0|\fat{y}_{\nmid \ell}^\prime)}\nonumber\\
\!\!\!\!&&\!\!\!\!\!\!\!\!=\log\frac{\textrm{Pr}(\beta^{(k)}_1=\lambda(s^{(k)}_\ell)^{-1}|\fat{y}_{\nmid \ell}^\prime)}{\textrm{Pr}(\beta^{(k)}_1=0|\fat{y}_{\nmid \ell}^\prime)}\nonumber\\
\!\!\!\!&&\!\!\!\!\!\!\!\!=\log\frac{\textrm{Pr}(\beta^{(k)}_1\fat{s}^{(k)}_{\nmid
\ell}=\lambda(s^{(k)}_\ell)^{-1}\fat{s}^{(k)}_{\nmid
\ell}|\fat{y}_{\nmid
\ell}^\prime)}{\textrm{Pr}(\beta^{(k)}_1\fat{s}^{(k)}_{\nmid
\ell}=\fat{0}_{L-1}|\fat{y}_{\nmid \ell}^\prime)}\nonumber\\
\!\!\!\!&&\!\!\!\!\!\!\!\!=\log\frac{\prod_{i=1,i\neq\ell}^L\textrm{Pr}(\gamma^{(k)}_{i}=\lambda(s^{(k)}_\ell)^{-1}s^{(k)}_i|\fat{y}^\prime_{i})}{\prod_{i=1,i\neq\ell}^L\textrm{Pr}(\gamma^{(k)}_{i}=0|\fat{y}^\prime_{i})}\nonumber\\
\!\!\!\!\!\!\!\!&&\!\!\!\!\!\!\!\!=\sum_{i=1,i\neq\ell}^LL^a_{\lambda(s^{(k)}_\ell)^{-1}s^{(k)}_i}(\gamma^{(k)}_{i}),\
\lambda=0,1,...,\alpha^{2^s-2} \label{eq:variable-node}
\end{eqnarray}
where $\fat{0}_{L-1}=(0,...,0)$ is a length-$(L-1)$ zero vector and
$(s^{(k)}_\ell)^{-1}$ is the inverse of $s^{(k)}_\ell$ in GF($2^s$).

The $\ell$-th, $1\leq \ell\leq L$, mapping node in $M_b$ transforms
the extrinsic symbol LLR vector $\fat{L}^e(\gamma^{(k)}_\ell)$ to
extrinsic chip LLRs of $c^{(k)}_{(\ell-1)s+n}$ as
\begin{eqnarray}
\!\!\!\!\!\!\!\!L^e(c^{(k)}_{(\ell-1)s+n})&&\!\!\!\!\!\!\!\!\!\!=\log\frac{\textrm{Pr}(c^{(k)}_{(\ell-1)s+n}=+1|\fat{y}_{\nmid \ell}^\prime)}{\textrm{Pr}(c^{(k)}_{(\ell-1)s+n}=-1|\fat{y}_{\nmid \ell}^\prime)}\nonumber\\
&&\!\!\!\!\!\!\!\!\!\!=\log\frac{\sum_{\lambda\in
\textrm{GF}(2^s),\Gamma_n^{-1}(\lambda)=+1}\textrm{Pr}(\gamma^{(k)}_\ell=\lambda|\fat{y}_{\nmid
\ell}^\prime)}{\sum_{\lambda\in
\textrm{GF}(2^s),\Gamma_n^{-1}(\lambda)=-1}\textrm{Pr}(\gamma^{(k)}_\ell=\lambda|\fat{y}_{\nmid
\ell}^\prime)}\nonumber\\
&&\!\!\!\!\!\!\!\!\!\!=\log\frac{\sum_{\lambda\in
\textrm{GF}(2^s)}\frac{1+\Gamma_n^{-1}(\lambda)}{2}\frac{\textrm{Pr}(\gamma^{(k)}_\ell=\lambda|\fat{y}_{\nmid
\ell}^\prime)}{\textrm{Pr}(\gamma^{(k)}_\ell=0|\fat{y}_{\nmid
\ell}^\prime)}}{\sum_{\lambda\in
\textrm{GF}(2^s)}\frac{1-\Gamma_n^{-1}(\lambda)}{2}\frac{\textrm{Pr}(\gamma^{(k)}_\ell=\lambda|\fat{y}_{\nmid
\ell}^\prime)}{\textrm{Pr}(\gamma^{(k)}_\ell=0|\fat{y}_{\nmid
\ell}^\prime)}}\nonumber\\
&&\!\!\!\!\!\!\!\!\!\!=\log\frac{\sum_{\lambda\in
\textrm{GF}(2^s)}(1\!+\!\Gamma_n^{-1}(\lambda))e^{L^e_\lambda(\gamma^{(k)}_\ell)}}{\sum_{\lambda\in
\textrm{GF}(2^s)}(1\!-\!\Gamma_n^{-1}(\lambda))e^{L^e_\lambda(\gamma^{(k)}_\ell)}},\
n=1,...,s.\label{eq:symbol-bit}
\end{eqnarray}
These extrinsic chip LLRs will be interleaved to update the priori
LLRs of the sum node decoding in (\ref{eq:ESE}).

\subsubsection{Hard Decision}\label{sec:decision}
The hard decision is performed on the output from the mapping node
in $M_a$ to the variable node in $U$. Using a similar principle as
in (\ref{eq:variable-node}), the first variable node in $V$
calculates a total posteriori LLR vector
$\fat{L}(\beta^{(k)}_1)\!=\!(L_{0}(\beta^{(k)}_1),L_{1}(\beta^{(k)}_1),...,L_{\alpha^{2^s-2}}(\beta^{(k)}_1))$
of $\beta^{(k)}_1$ as
\begin{equation}\label{eq:symboldecision}
L_\lambda(\beta^{(k)}_1)=\sum_{i=1}^LL^a_{\lambda
s^{(k)}_i}(\gamma^{(k)}_{i}),\ \lambda=0,1,...,\alpha^{2^s-2}.
\end{equation}
The first mapping node in $M_a$ transforms this field symbol LLR
vector to $s$ bit LLRs using a similar principle as in
(\ref{eq:symbol-bit})
\begin{equation}\label{eq:bitdecision}
L(u_n)=\log\frac{\sum_{\lambda\in
\textrm{GF}(2^s)}(1\!+\!\Gamma_n^{-1}(\lambda))e^{L_\lambda(\beta^{(k)}_1)}}{\sum_{\lambda\in
\textrm{GF}(2^s)}(1\!-\!\Gamma_n^{-1}(\lambda))e^{L_\lambda(\beta^{(k)}_1)}},\
n=1,...,s
\end{equation}
for bit decision. Note that the mapping node in $M_a$ does not
provide any message to the variable node in $V$ during the iterative
decoding. Here we give a bit decision algorithm, in which the hard
decision is performed on the LLRs of information bits in
(\ref{eq:bitdecision}). The hard decision can also be performed on
the field symbol LLR in (\ref{eq:symboldecision}) to obtain an
estimation of $\beta^{(k)}_1$.

\section{Analysis of EXIT Function of FF-DES}\label{sec:analysis}
To show the performance gain of joint spreading quantitatively, we
analyze the EXIT function of the FF-DES. In
Section~\ref{sec:appEXIT}, we give an approximate EXIT function and
show that this EXIT function asymptotically approaches a line. In
Section~\ref{sec:slope}, we derive the asymptotic slope of the EXIT
function.

\subsection{Approximate EXIT Function}\label{sec:appEXIT}
The EXIT function describes a relation between the priori input and
the extrinsic output of a decoding. Both the input and the output of
the decoding are measured by mutual information or LLR mean value
based on the Gaussian approximation
\cite{gaussian,Convergence,modern,lin}. In this section, we give an
approximate EXIT function of the FF-DES using the measure of LLR
mean value.

The EXIT function of FF-DES describes the relation between the mean
value of a priori chip LLR and that of the extrinsic chip LLR.
Generally, this EXIT function is determined by the realizations of a
specific chip, mapping $\Gamma$, and spreading vector
$\fat{s}^{(k)}$. Here we consider random chip, random mapping
$\Gamma$, and random spreading vector $\fat{s}^{(k)}$, all of which
are uniformly generated from all their possible realizations. We
derive an expected EXIT function that is an average for all the
possible chip, mapping, and spreading vector realizations.

Since as stated in Section~\ref{sec:decoding}, the decoding
operation for each chip of each user is the same, we omit
superscript $(k)$ to consider the chip $c_{(\ell-1)s+n},
1\leq\ell\leq L, 1\leq n\leq s$, in our analysis. The EXIT function
is the function between
$E[c_{(\ell\!-\!1)s+n}L^e(c_{(\ell\!-\!1)s+n})]$ and
$E[c_{(i-1)s+m}L^a(c_{(i-1)s+m})],i\neq \ell$, $1\leq m\leq s$,
where $E[\cdot]$ takes the expectation of a random variable.

Combining (\ref{eq:bit-symbol}), (\ref{eq:variable-node}), and
(\ref{eq:symbol-bit}) in the FF-DES, we write $L^e(c_{(\ell-1)s+n})$
as a function of $L^a(c_{(i-1)s+m})$, $i\neq\ell, 1\leq m\leq s$,
\begin{eqnarray}
L^e(c_{(\ell-1)s+n})\!\!\!\!\!\!\!\!&&=\log\frac{\sum_{\lambda\in
\textrm{GF}(2^s)}(1\!+\!\Gamma_n^{-1}(\lambda))e^{\sum_{i=1,i\neq\ell}^L\sum_{m=1}^s\!\frac{\Gamma^{-1}_m(\lambda(s_\ell)^{-1}s_i)-\Gamma^{-1}_m(0)}{2}L^a(c_{(i-1)s+m})}}{\sum_{\lambda\in
\textrm{GF}(2^s)}(1\!-\!\Gamma_n^{-1}(\lambda))e^{\sum_{i=1,i\neq\ell}^L\sum_{m=1}^s\!\frac{\Gamma^{-1}_m(\lambda(s_\ell)^{-1}s_i)-\Gamma^{-1}_m(0)}{2}L^a(c_{(i-1)s+m})}}\nonumber\\
&&=\log\frac{\sum_{\lambda\in
\textrm{GF}(2^s)}(1\!+\!\Gamma_n^{-1}(\lambda))e^{\rho(\lambda)}}{\sum_{\lambda\in
\textrm{GF}(2^s)}(1\!-\!\Gamma_n^{-1}(\lambda))e^{\rho(\lambda)}}\label{eq:DES}
\end{eqnarray}
where
$\rho(\lambda)\triangleq\frac{1}{2}\sum_{i=1,i\neq\ell}^L\sum_{m=1}^s\!\Gamma^{-1}_m(\lambda(s_\ell)^{-1}s_i)L^a(c_{(i-1)s+m})$.

By the Gaussian approximation \cite{gaussian}\cite{Convergence},
$c_{(i-1)s+m}L^a(c_{(i-1)s+m})$ $\sim\mathcal {N}(m_a,2m_a),1\leq
i\leq L, 1\leq m\leq s$, i.i.d., and
$c_{(\ell-1)s+n}L^e(c_{(\ell-1)s+n})\sim\mathcal {N}(m_e,2m_e)$,
where $\mathcal {N}(\mu,\sigma^2)$ denotes the Gaussian distribution
with mean $\mu$ and variance $\sigma^2$. Based on (\ref{eq:DES}),
the EXIT function becomes
\begin{eqnarray}
\ \ \ \ \ \ \ \ \ \ \ \ m_e\!\!\!\!\!\!\!\!&&=E[c_{(\ell-1)s+n}L^e(c_{(\ell-1)s+n})]\nonumber\\
\ \ \ \ \ \ \ \ \ \ \ \
&&=E[c_{(\ell-1)s+n}\log\frac{\sum_{\lambda\in
\textrm{GF}(2^s)}(1\!+\!\Gamma_n^{-1}(\lambda))e^{\rho(\lambda)}}{\sum_{\lambda\in
\textrm{GF}(2^s)}(1\!-\!\Gamma_n^{-1}(\lambda))e^{\rho(\lambda)}}]\nonumber\\
\ \ \ \ \ \ \ \ \ \ \ \ &&=E[\log\frac{\sum_{\lambda\in
\textrm{GF}(2^s)}(1\!+c_{(\ell-1)s+n}\Gamma_n^{-1}(\lambda))e^{\rho(\lambda)}}{\sum_{\lambda\in
\textrm{GF}(2^s)}(1\!-c_{(\ell-1)s+n}\Gamma_n^{-1}(\lambda))e^{\rho(\lambda)}}]\nonumber\\
\ \ \ \ \ \ \ \ \ \ \ \
&&=E[\log\frac{(1\!+c_{(\ell-1)s+n}\Gamma_n^{-1}(\gamma_{\ell}))e^{\rho(\gamma_\ell)}+\sum_{\lambda\in
\textrm{GF}(2^s),\lambda\neq\gamma_{\ell}}(1\!+c_{(\ell-1)s+n}\Gamma_n^{-1}(\lambda))e^{\rho(\lambda)}}{(1\!-c_{(\ell-1)s+n}\Gamma_n^{-1}(\gamma_{\ell}))e^{\rho(\gamma_\ell)}+\sum_{\lambda\in
\textrm{GF}(2^s),\lambda\neq\gamma_{\ell}}(1\!-c_{(\ell-1)s+n}\Gamma_n^{-1}(\lambda))e^{\rho(\lambda)}}]\nonumber\\
\ \ \ \ \ \ \ \ \ \ \ \
&&=E[\log\frac{2e^{\rho(\gamma_\ell)}\!+\!\sum_{\lambda\in
\textrm{GF}(2^s),\lambda\neq\gamma_{\ell}}(1\!+\!c_{(\ell-1)s+n}\Gamma_n^{-1}(\lambda))e^{\rho(\lambda)}}{\sum_{\lambda\in
\textrm{GF}(2^s),\lambda\neq\gamma_{\ell}}(1\!-c_{(\ell-1)s+n}\Gamma_n^{-1}(\lambda))e^{\rho(\lambda)}}]\label{eq:exit1}\\
\ \ \ \ \ \ \ \
&&=E[2\rho(\gamma_\ell)]+E[\log\frac{1+\frac{1}{2}\sum_{\lambda\in
\textrm{GF}(2^s),\lambda\neq\gamma_{\ell}}(1\!+\!c_{(\ell-1)s+n}\Gamma_n^{-1}(\lambda))e^{\rho(\lambda)-\rho(\gamma_\ell)}}{\frac{1}{2}\sum_{\lambda\in
\textrm{GF}(2^s),\lambda\neq\gamma_{\ell}}(1\!-c_{(\ell-1)s+n}\Gamma_n^{-1}(\lambda))e^{\rho(\lambda)+\rho(\gamma_\ell)}}]\nonumber\\
\ \ \ \ \ \ \ \ \ \ \ \
&&=s(L-1)m_a-E[\log(\sum_{\lambda\in\lambda^-}e^{\rho(\gamma_\ell)+\rho(\lambda)})]+E[\log(1+\sum_{\lambda\in\lambda^+}e^{-(\rho(\gamma_\ell)-\rho(\lambda))})].\label{eq:exit}
\end{eqnarray}
Eq.~(\ref{eq:exit1}) is from the fact
$\Gamma_n^{-1}(\gamma_{\ell})=c_{(\ell-1)s+n}$ since $\gamma_\ell$
is the correct symbol. In (\ref{eq:exit}), the first term is from
$\rho(\gamma_\ell)=\frac{1}{2}\sum_{i=1,i\neq\ell}^L\sum_{m=1}^sc_{(i-1)s+m}L^a(c_{(i-1)s+m})$
due to $\Gamma^{-1}_m(\gamma_{\ell}(s_\ell)^{-1}s_i)=c_{(i-1)s+m}$.
In the second and third terms,
$\lambda^-\triangleq\{\lambda|\lambda\in \textrm{GF}(2^s),
\Gamma_n^{-1}(\lambda)=-c_{(\ell-1)s+n}\}$ and
$\lambda^+\triangleq\{\lambda|\lambda\in \textrm{GF}(2^s),
\lambda\neq\gamma_\ell, \Gamma_n^{-1}(\lambda)=c_{(\ell-1)s+n}\}$.
It holds that $|\lambda^-|=2^{s-1}$, $|\lambda^+|=2^{s-1}-1$, where
$|\cdot|$ takes the cardinality of a set.

Now we give an approximation for the EXIT function in
(\ref{eq:exit}). For $\lambda\in\lambda^-$, we have
\begin{eqnarray}
\rho(\gamma_\ell)+\rho(\lambda)&&\!\!\!\!\!\!\!\!=\frac{1}{2}\sum_{i=1,i\neq\ell}^L\sum_{m=1}^s(c_{(i-1)s+m}+\Gamma^{-1}_m(\lambda(s_\ell)^{-1}s_i))L^a(c_{(i-1)s+m})\nonumber\\
&&\!\!\!\!\!\!\!\!
=\sum_{i=1,i\neq\ell}^L\sum_{m=1}^s\frac{(1+c_{(i-1)s+m}\Gamma^{-1}_m(\lambda(s_\ell)^{-1}s_i))}{2}c_{(i-1)s+m}L^a(c_{(i-1)s+m}).\label{eq:rho-}
\end{eqnarray}
Similarly, for $\lambda\in\lambda^+$, we have
\begin{eqnarray}
\ \ \
\rho(\gamma_\ell)-\rho(\lambda)=\sum_{i=1,i\neq\ell}^L\sum_{m=1}^s\frac{(1\!-\!c_{(i-1)s+m}\Gamma^{-1}_m(\lambda(s_\ell)^{-1}\!s_i))}{2}
c_{(i-1)s+m}L^a(c_{(i-1)s+m}).\label{eq:rho+}
\end{eqnarray}

Vector $(c_{(i-1)s+1}\Gamma^{-1}_1(\lambda(s_\ell)^{-1}s_i),
...,c_{is}\Gamma^{-1}_s(\lambda(s_\ell)^{-1}s_i))$ in
(\ref{eq:rho-}) and (\ref{eq:rho+}) is a bit-wise correlation
between the binary vectors for the correct symbol and an error
symbol. Due to random chip $c_{(i-1)s+m}$, random mapping $\Gamma$,
and random spreading element $s_i$, this bit-wise correlation vector
is approximately uniformly distributed on the set that includes all
the binary vector in $\mathcal {X}^s$ except all-one vector
$\textbf{1}_s=(1,...,1)$. Let set $\Omega_s^-$ include all the
binary-$\{0,1\}$ vectors of length $s$ except $\textbf{1}_s$, and
$\Omega_s^+$ include all the binary-$\{0,1\}$ vectors of length $s$
except $\textbf{0}_s$. Let $\mathcal {U}(\Omega)$ denote the uniform
distribution on set $\Omega$. We have the following approximation of
the EXIT function (\ref{eq:exit})
\begin{eqnarray}
m_e\approx
s(L-1)m_a-E[\log(\sum_{j=1}^{2^{s-1}}e^{\sum_{i=1}^{L-1}\fat{r}_{j,i}\fat{\hbar}_i^\textrm{T}})]+E[\log(1+\sum_{j=1}^{2^{s-1}-1}e^{-\sum_{i=1}^{L-1}\fat{r}_{j,i}^\prime\fat{\hbar}_i^\textrm{T}})]\triangleq
\varphi(m_a)\label{eq:exitapp}
\end{eqnarray}
where $\fat{r}_{j,i},j=1,...,2^{s-1},i=1,...,L-1$, are i.i.d. with
$\fat{r}_{j,i}\!\sim\! \mathcal {U}(\Omega_s^-)$. Vectors
$\fat{r}_{j,i}^\prime,j\!=\!1,...,2^{s-1}\!-1$, $i=1,...,L-1$, are
i.i.d. with $\fat{r}_{j,i}^\prime\sim \mathcal {U}(\Omega_s^+)$.
Elements $h_{i,m},i=1,...,L,m=1,...,s$, in
$\fat{\hbar}_i,i=1,...,L$, are all i.i.d. with $h_{i,m}\sim \mathcal
{N}(m_a,2m_a)$.

\begin{figure}
\includegraphics[width=3.3 in]{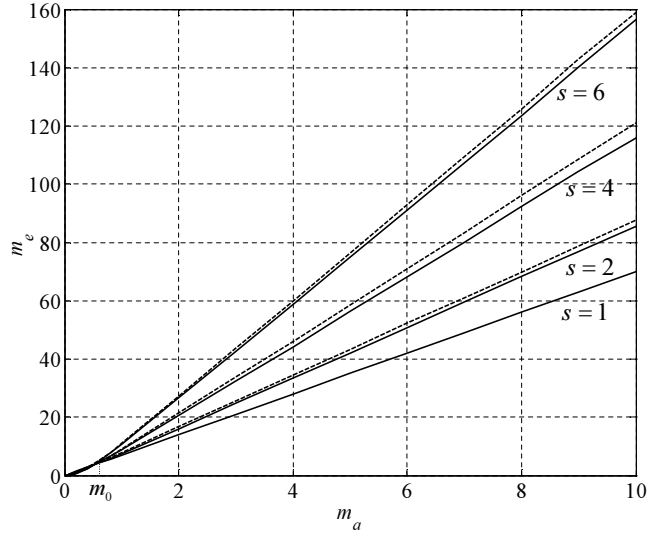}
\centering
%
\caption{EXIT functions (solid lines) and their approximations of
$\varphi(m_a)$ (dashed lines) of the FF-DES for $L=8$ and
$s=1,2,4,6$.} \label{fig:exit8}
\end{figure}
\begin{figure}
\includegraphics[width=3.3 in]{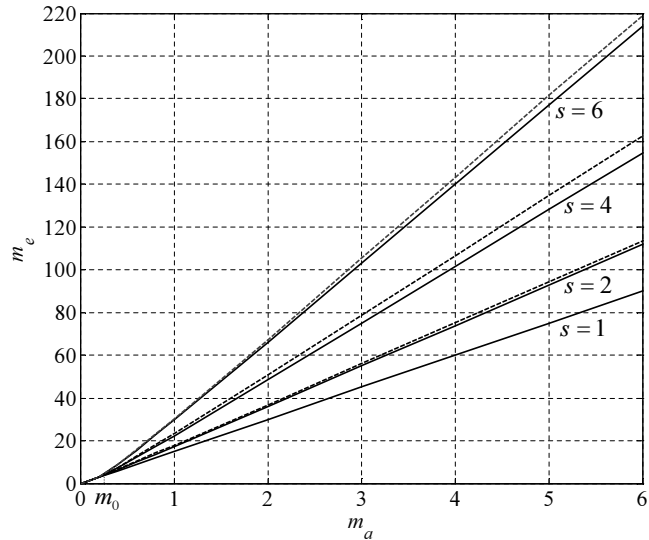}
\centering
%
\caption{EXIT functions (solid lines) and their approximations of
$\varphi(m_a)$ (dashed lines) of the FF-DES for $L=16$ and
$s=1,2,4,6$.} \label{fig:exit16}
\end{figure}

To see the accuracy of the approximation in (\ref{eq:exitapp}),
Figs.~\ref{fig:exit8} and \ref{fig:exit16} illustrate the EXIT
function in (\ref{eq:exit}) and $\varphi(m_a)$ in (\ref{eq:exitapp})
for $L=8,16$ and $s=1,2,4,6$ by Monte Carlo simulations. We see that
$\varphi(m_a)$ is in fact a tight upper bound of the EXIT function.
The ratio between $\varphi(m_a)$ and the EXIT function is less than
1.06 for $m_a\leq 10$ in both figures. Based on this fact, in the
rest analysis of this paper, we employ approximate EXIT function
$\varphi(m_a)$ of the FF-DES.

Moreover, we have the following observation on the approximate EXIT
function. There exists a cutoff point $m_0$ usually less than 1. For
$m_a\geq m_0$, every curve of the approximate EXIT function in
Figs.~\ref{fig:exit8} and \ref{fig:exit16} asymptotically approaches
a line. The asymptotic slope of the approximate EXIT function
increases as $s$ increases. This means that by joint spreading for
$s$ information bits, an SINR gain is obtained with respect to that
of $s=1$. This SINR gain will be enhanced as length $s$ increases.
Furthermore, by comparing Fig.~\ref{fig:exit8} and
Fig.~\ref{fig:exit16} we can see that given $s$, the asymptotic
slope of the approximate EXIT function for $L=16$ is larger than
that for $L=8$. This is due to that as spreading length $L$
increases, the advantage of joint spreading is enlarged.

\begin{remark}
Although we have focused on the asymptotic slope of the EXIT
function of the FF-DES when $m_a$ is large above, the EXIT function
at $m_a$ near 0 is also noteworthy. If the number of users $K$ is
large, due to a large multi-user interference, at the beginning of
decoding iteration the extrinsic output of the ESE is always very
small. This requires the EXIT function of the FF-DES to be
sufficiently large at $m_a$ near 0, or else the iterative decoding
would fail at the beginning of iteration. As revealed in
\cite{supermap} and \cite{supermapj}, conventional spreading scheme
is widely used in the MAC since it can provide a large EXIT function
at $m_a$ near 0. We see that in Figs.~\ref{fig:exit8} and
\ref{fig:exit16}, the finite field spreading can provide a similar
EXIT function as that of the conventional spreading scheme ($s=1$)
at $m_a<m_0$, and can also work in the environment of large
multi-user interference.
\end{remark}


\subsection{Asymptotic Slope}\label{sec:slope}
Section~\ref{sec:appEXIT} has shown that when $m_a$ is larger than
cutoff point $m_0$, the approximate EXIT function of FF-DES
asymptotically approaches a line. In this section, we analyze this
slope by deriving the asymptotic differential of $\varphi(m_a)$ in
(\ref{eq:exitapp}).

\begin{theorem}\label{thm:diversity}
Let $g(s,L)$ be a function of $s$ and $L$. Let $m_0\geq0$ be a
constant. Suppose that $\varphi (m_a)=\varphi(m_0)+g(s,L)(m_a-m_0)$
holds for $m_a\geq m_0$, we have
\begin{equation}
g(s,L)=L-1+\frac{1}{(2^s-1)^{(L-1)2^{s-1}}}\sum_{k=1}^{(s-1)(L-1)}\left(\sum_{(n_1,...,n_{L-1})\in
\theta(k-1)}\prod_{i=1}^{L-1}\binom{s}{n_i}\right)^{2^{s-1}}\label{eq:diversity}
\end{equation}
where $\theta(k\!-\!1)\triangleq
\{n_1,...,n_{L-1}|\sum_{i=1}^{L-1}n_i\leq k\!-\!1,0\leq n_i\leq
s\!-\!1, n_i\in Z\}$ with integer set $Z$. \myQED
\end{theorem}

To prove Theorem~\ref{thm:diversity}, we first prove the following
lemma.
\begin{lemma}\label{lem:lim}
Let $a\geq1$ be a constant. Let
$\widehat{\fat{r}}_j=[\widehat{r}_{j,1},...,\widehat{r}_{j,m}],\widehat{r}_{j,i}\in\{-1,0,1\},j=1,...,n$,
be i.i.d. random vectors with $w(\widehat{\fat{r}}_j)\triangleq
\sum_{i=1}^m\widehat{r}_{j,i}\leq 0$, and
$w^+(\widehat{\fat{r}}_j)\triangleq
\sum_{i=1}^m|\widehat{r}_{j,i}|>0$ for all $j$. Let
$\widehat{\fat{\hbar}}$ be a length-$m$ vector whose elements are
i.i.d. with Gaussian distribution $\mathcal {N}(\mu,2\mu),\mu>0$. It
holds that
\begin{equation}
\lim_{\mu\rightarrow
+\infty}\frac{E[\log(a+\sum_{j=1}^ne^{\widehat{\fat{r}}_j\widehat{\fat{\hbar}}^\textrm{T}})]}{\mu}=0.\label{eq:lemma}
\end{equation}\myQED
\end{lemma}

\emph{Proof:} We first have
\begin{eqnarray}
\!\!0\leq
E[\log(a+\sum_{j=1}^ne^{\widehat{\fat{r}}_j\widehat{\fat{\hbar}}^\textrm{T}})]\!\!\!\!\!\!\!&&\leq
E[\log(a+\sum_{j=1}^ne^{\widehat{\fat{r}}_j\widehat{\fat{\hbar}}^\textrm{T}-w(\widehat{\fat{r}}_j)\mu})]
\nonumber\\
&&\leq
E[\log((n+1)\max\{a,e^{|\widehat{\fat{r}}_1\widehat{\fat{\hbar}}^\textrm{T}-w(\widehat{\fat{r}}_1)\mu|},...,e^{|\widehat{\fat{r}}_n\widehat{\fat{\hbar}}^\textrm{T}-w(\widehat{\fat{r}}_n)\mu|}\})]\nonumber\\
&&\leq \log(n+1)+E[\log
a+\sum_{j=1}^n|\widehat{\fat{r}}_j\widehat{\fat{\hbar}}^\textrm{T}-w(\widehat{\fat{r}}_j)\mu|]\nonumber\\
&&=\log(a(n+1))\!+\!n\sum_{k=1}^mE[|\widehat{\fat{r}}_1\widehat{\fat{\hbar}}^\textrm{T}\!\!-\!w(\widehat{\fat{r}}_1)\mu|\ |w^+(\widehat{\fat{r}}_1)\!=\!k]\textrm{Pr}(w^+(\widehat{\fat{r}}_1)\!=\!k)\nonumber\\
&&=\log(a(n+1))\!+\!2n\sum_{k=1}^m\sqrt{\frac{k\mu}{\pi}}\textrm{Pr}(w^+(\widehat{\fat{r}}_1)\!=\!k) \label{eq:jifen}\\
&&\leq\log(a(n+1))\!+\!2n\sqrt{\frac{m\mu}{\pi}} \nonumber
\end{eqnarray}
where (\ref{eq:jifen}) is due to that under the condition
$w^+(\widehat{\fat{r}}_1)\!=\!k$,
$(\widehat{\fat{r}}_1\widehat{\fat{\hbar}}^\textrm{T}\!\!-\!w(\widehat{\fat{r}}_1)\mu)\sim
\mathcal {N}(0,2k\mu)$ holds for each realization of
$\widehat{\fat{r}}_1$. Thus,
\begin{eqnarray}
0\leq\lim_{\mu\rightarrow
+\infty}\frac{E[\log(a+\sum_{j=1}^ne^{\widehat{\fat{r}}_j\widehat{\fat{\hbar}}^\textrm{T}})]}{\mu}\leq\lim_{\mu\rightarrow
+\infty}\frac{\log(a(n+1))\!+\!2n\sqrt{\frac{m\mu}{\pi}}}{\mu}
=0.\nonumber
\end{eqnarray}
The lemma is proved.

\emph{Proof of Theorem~\ref{thm:diversity}:} Let
$\jmath_{\max}\triangleq\{j^*|j^*\!=\!\arg\max_j\{\sum_{i=1}^{L-1}w(\fat{r}_{j,i})\}$
be the set of subscripts that maximize the Hamming weight of joint
vector $(\fat{r}_{j,1},...,\fat{r}_{j,L-1})$ in (\ref{eq:exitapp}).
Given an element of $j^*\in\jmath_{\max}$, we have set
$\jmath^*\triangleq\{j|\fat{r}_{j,i}=\fat{r}_{j^*,i},i=1,...,L-1\}\subseteq\jmath_{\max}$.

Since $\varphi(m_a)=\varphi(m_0)+g(s,L)(m_a-m_0)$ holds as
$m_a\rightarrow+\infty$ by assumption, using (\ref{eq:exitapp}) we
have
\begin{eqnarray}
g(s,L)\!\!\!\!\!\!\!\!&&=\lim_{m_a\rightarrow+\infty}\frac{\varphi(m_a)-\varphi(m_0)}{m_a-m_0}=\lim_{m_a\rightarrow+\infty}\frac{\varphi(m_a)}{m_a}\nonumber\\
&&=
s(L-1)-\lim_{m_a\rightarrow+\infty}\frac{E[\log(\sum_{j=1}^{2^{s-1}}e^{\sum_{i=1}^{L-1}\fat{r}_{j,i}\fat{\hbar}_i^\textrm{T}})]}{m_a}+\lim_{m_a\rightarrow+\infty}\frac{E[\log(1\!+\!\sum_{j=1}^{2^{s-1}-1}e^{-\sum_{i=1}^{L-1}\fat{r}_{j,i}^\prime\fat{\hbar}_i^\textrm{T}})]}{m_a}\nonumber\\
&&=s(L-1)-\lim_{m_a\rightarrow+\infty}\frac{E[\log(\sum_{j=1}^{2^{s-1}}e^{\sum_{i=1}^{L-1}\fat{r}_{j,i}\fat{\hbar}_i^\textrm{T}})]}{m_a}\label{eq:0lim1}\\
&&
=s(L-1)-\lim_{m_a\rightarrow+\infty}\frac{E[\sum_{i=1}^{L-1}\fat{r}_{j^*,i}\fat{\hbar}_i^\textrm{T}]}{m_a}-\lim_{m_a\rightarrow+\infty}\frac{E[\log(|\jmath^*|+\sum_{j=1,j\not\in
\jmath^*}^{2^{s-1}}e^{-\sum_{i=1}^{L-1}(\fat{r}_{j^*,i}-\fat{r}_{j,i})\fat{\hbar}_i^\textrm{T}})]}{m_a}\nonumber\\
&&=s(L-1)-E[\sum_{i=1}^{L-1}w(\fat{r}_{j^*,i})]\label{eq:0lim2}\\
&&=s(L-1)-\sum_{k=1}^{(s-1)(L-1)}k\textrm{Pr}(\sum_{i=1}^{L-1}w(\fat{r}_{j^*,i})=k)\nonumber\\
&&=
s(L-1)-\sum_{k=1}^{(s-1)(L-1)}k(\textrm{Pr}(\sum_{i=1}^{L-1}w(\fat{r}_{j^*,i})\leq k)-\textrm{Pr}(\sum_{i=1}^{L-1}w(\fat{r}_{j^*,i})\leq k-1))\nonumber\\
&&=s(L\!-\!1)-(s\!-\!1)(L\!-\!1)\textrm{Pr}(\sum_{i=1}^{L\!-\!1}w(\fat{r}_{j^*,i})\leq
(s\!-\!1)(L\!-\!1))+\sum_{k=1}^{(s\!-\!1)(L\!-\!1)}\textrm{Pr}(\sum_{i=1}^{L\!-\!1}w(\fat{r}_{j^*,i})\leq
k\!-\!1)\nonumber\\
&&=L-1+\sum_{k=1}^{(s-1)(L-1)}\textrm{Pr}(\sum_{i=1}^{L-1}w(\fat{r}_{1,i})\leq
k-1,...,\sum_{i=1}^{L-1}w(\fat{r}_{2^{s-1},i})\leq
k-1)\nonumber\\
&&=L-1+\sum_{k=1}^{(s-1)(L-1)}\left(\textrm{Pr}(\sum_{i=1}^{L-1}w(\fat{r}_{1,i})\leq
k-1)\right)^{2^{s-1}}\label{eq:iid}\\
&&=L-1+\frac{1}{(2^s-1)^{(L-1)2^{s-1}}}\sum_{k=1}^{(s-1)(L-1)}\left(\sum_{(n_1,...,n_{L-1})\in
\theta(k-1)}\prod_{i=1}^{L-1}\binom{s}{n_i}\right)^{2^{s-1}}.\label{eq:diversityproof}
\end{eqnarray}
Eq.~(\ref{eq:0lim1}) is from Lemma~\ref{lem:lim}.
Eq.~(\ref{eq:0lim2}) is due to that elements of $\fat{\hbar}_i$ are
i.i.d. with distribution $\mathcal {N}(m_a,2m_a)$ and
Lemma~\ref{lem:lim}. Eq.~(\ref{eq:iid}) is due to that
$\sum_{i=1}^{L-1}w(\fat{r}_{j,i}),j=1,...,2^{s-1}$, are i.i.d..
Eq.~(\ref{eq:diversityproof}) is due to that
$\fat{r}_{1,i},i=1,...,L-1$, are i.i.d. with $\fat{r}_{1,i}\sim
\mathcal {U}(\Omega_s^-)$. The theorem is proved.\myQED

For $s=1$, $g(1,L)=L-1$ is the slope of the EXIT function of the DES
for the conventional spreading scheme of IDMA
\cite{IDMA,idmaj,analysisIDMA}. The last term in
(\ref{eq:diversityproof}) is the slope gain from the joint
spreading.

\section{Slope of BER}\label{sec:ber}
In this section, we show that the asymptotic slope of approximate
EXIT function analyzed in Section~\ref{sec:slope} in fact is the
absolute slope of BER curve at the low BER region. We verify our
analysis by BER Monte Carlo simulations for practical finite field
multiple-access systems.

Let $\phi(m_a,\frac{E_b}{N_0})$ be the EXIT function of the ESE
(Appendix~\ref{app:ese}). The mean of extrinsic LLR converges to
infinity as the number of iteration increases if and only if
$\phi(x,\frac{E_b}{N_0})>\varphi^{-1}(x)$ holds for $x\geq 0$
\cite{gaussian,Convergence,modern,lin}, where $\varphi^{-1}(\cdot)$
is the inverse function of $\varphi(\cdot)$ in (\ref{eq:exitapp}).
Since $\phi(x,\frac{E_b}{N_0})\leq 4\frac{E_b}{LN_0}$
(Lemma~\ref{lem:ESElim} in Appendix~\ref{app:ese}) and
$\lim_{x\rightarrow+\infty} \varphi^{-1}(x)=\!+\infty$, to converge
to infinite mean of LLR, $\frac{E_b}{N_0}$ should approach infinity.
This is due to that no channel code is employed for each user.

Since for a given $x$, $\varphi^{-1}(x)$ is a decreasing function of
spreading length $L$, $\phi(x,\frac{E_b}{N_0})>\varphi^{-1}(x)$
holds for $x\geq 0$ as $L$ is sufficiently large and
$\frac{E_b}{N_0}\rightarrow+\infty$. In this case, the mean of the
extrinsic LLR converges to infinity as the number of iteration
increases. Due to $\lim_{x\rightarrow
+\infty}\phi(x,\frac{E_b}{N_0})=4\frac{E_b}{LN_0}$ from
Lemma~\ref{lem:ESElim}, the asymptotic BER of hard decision at the
information node in $U$ is estimated as
\cite{gaussian,Convergence,modern,lin}
\begin{equation}\label{eq:ber}
P_e\approx Q(\sqrt{2g(s,L+1)\frac{E_b}{LN_0}}) <
e^{-\widetilde{g}(s,L)\frac{E_b}{N_0}}
\end{equation}
where we have used $\varphi(x)\approx g(s,L)x$ as $x$ is large,
$Q(x)<e^{-x^2/2}$, and $\widetilde{g}(s,L)\triangleq g(s,L+1)/L$ is
referred to as a standard slope. Note that we have used slope
$g(s,L+1)$ instead of $g(s,L)$ in (\ref{eq:ber}) since the hard
decision is performed based on the total LLR in
(\ref{eq:bitdecision}) other than the extrinsic LLR in
(\ref{eq:symbol-bit}).

\begin{figure}
\includegraphics[width=2.6 in]{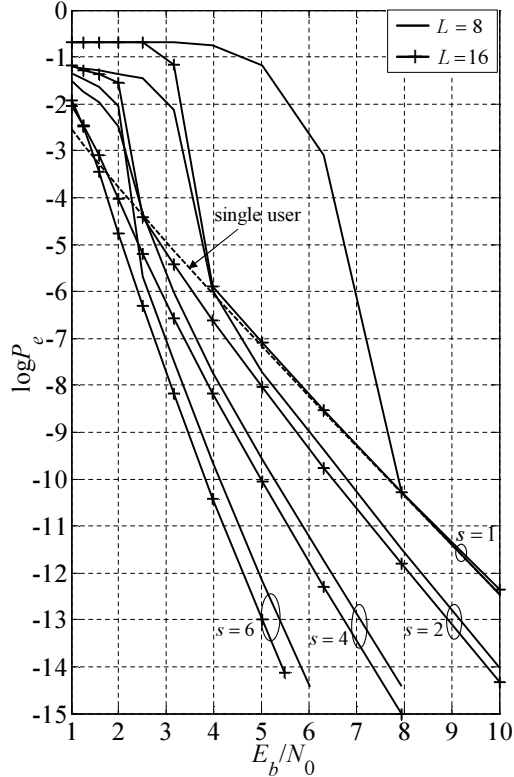}
\centering
%
\vspace{-2mm} \caption{BER curves of ($K=8$)-user finite field
spreading multiple-access systems with $L=8,16$ and $s=1,2,4,6$. The
information vector length is $sN=12000$ and the number of decoding
iteration is 50.} \label{fig:ber}
\end{figure}

In fact as the BER approaches 0, the upper bound in (\ref{eq:ber})
is very tight. Eq.~(\ref{eq:ber}) indicates that at the low BER
region, the BER curve is approximate to a line with absolute slope
$\widetilde{g}(s,L)$. Since $\widetilde{g}(1,L)=1$ holds for an
arbitrary $L$, at the low BER region, the BER curve of conventional
spreading scheme $(s=1)$ always converges to that of (single-user)
uncoded BPSK transmission. Unfortunately, this predicament can not
be improved by increasing spreading length $L$. In our work, since
$\widetilde{g}(s,L)>1$ holds for $s,L\geq2$, the BER curve of finite
field spreading with $s,L\geq2$ has a larger absolute slope than
that of single-user transmission with BPSK modulation, and the
absolute slope of the BER curve increases as $s$ increases. In
addition, there is a special case of $L=1$, i.e., code rate of each
user is 1. It is easy to show that in this case, the BER can never
approach 0 for the number of users $K\geq2$ due to code rate 1 of
each user. For single-user ($K=1$) finite field spreading
multiple-access, since $\widetilde{g}(s,1)\approx 1$ for an
arbitrary $s$, at the low BER region, the absolute slope of the BER
curve approaches that of single-user transmission with BPSK
modulation.

We verify our analysis above, obtained based on the Gaussian
approximation and the approximation of (\ref{eq:exitapp}), by BER
Monte Carlo simulations on practical systems. We consider
$(K\!=\!8)$-user finite field spreading multiple-access system with
$L=8,16$ and $s=1,2,4,6$. Fig.~\ref{fig:ber} illustrates BER curves
obtained by Monte Carlo simulations. Since the EXIT function theory
is based on the assumption of infinite code length
\cite{gaussian,Convergence,modern,lin}, in our simulations we employ
a large code length with the information vector length of
$sN=12000$. The number of decoding iteration is 50. Both mapping
$\Gamma(\cdot)$ and the chip-level interleaving are random, and the
BER curve illustrates an average BER for all the possible mapping
and interleaving realizations. We use horizontal coordinate
$\frac{E_b}{N_0}$ and vertical coordinate $\log P_e$ (natural
logarithm). We see that at the low BER region, all BER curves are
approximate to lines. We compare the absolute slopes of BER curves
at low BER region in Fig.~\ref{fig:ber} with standard slope
$\widetilde{g}(s,L)$ obtained by Theorem~\ref{thm:diversity} in
Table~\ref{tab:berslope}. The difference is less than $0.1$ for all
the comparison pairs. The two curves for $s=1$ overlap with each
other and converge to the BER curve of single-user transmission with
BPSK modulation. Curves for $s\geq 2$ have larger absolute slopes
than that of single-user transmission with BPSK modulation. All
these phenomenons coincide with our analysis.

\begin{table}
\caption{Comparison between absolute slopes of BER curves at low BER
region in Fig.~\ref{fig:ber} and $\widetilde{g}(s,L)$}\vspace{-4mm}
\label{tab:berslope}
\begin{center}
\begin{tabular}{|c|c|c|c|c|c|c|c|c|}

\hline  \multicolumn{1}{|c|}{$L$} & \multicolumn{4}{|c|}{8}& \multicolumn{4}{|c|}{16} \\
\hline
$s$ &1&2&4&6&1&2&4&6 \\
\hline
BER slope &1.0625&1.2262&1.6364&2.2352&1.0028&1.2373& 1.7269&2.3543 \\
\hline
$\widetilde{g}(s,L)$ &1&1.2411&1.7002&2.2095&1&1.2675&1.8240&2.4493 \\
\hline
\end{tabular}
\end{center}
\end{table}

\section{Conclusions}\label{sec:conclude}
In this paper, we proposed a finite field spreading scheme for the
MAC, in which the spreading for multiple bits is performed jointly
by the finite field multiplication. Under the iterative decoding, an
SINR gain is obtained during the joint spreading. The multi-user
finite field spreading multiple-access achieves a lower BER than
that of the single-user transmission with BPSK modulation under the
iterative decoding.

We considered the finite field spreading multiple-access system
without channel coding for each user. If there is a channel code
employed as an outer code for each user, our EXIT function analysis
of the FF-DES is also available from the theory of concatenated code
\cite{Convergence,modern,lin}.

There are still a number of problems in connection with our work
that seem to deserve further investigation. In our analysis, we
considered an average EXIT function and asymptotic slope for all the
mapping and interleaving. In fact, different mapping and
interleaving have different EXIT functions that provide different
BER performance. Design of spreading and interleaving to achieve
better BER performance is an interesting future work.

In the field field spreading of Fig.~\ref{fig:factorfinite}, every
$s$ information bits are spread into $L$ length-$s$ vectors, each of
which can be regarded as a codeword of a rate-1 code. These $L$
rate-1 codes constitute a finite field spreading code with rate
$\frac{1}{L}$. The $L$ multiplications on finite field can be
regarded as $L$ rate-1 encoding operations that determine the $L$
rate-1 codes. These rate-1 encoding can also realized by other
methods, or we can employ codes with rate less than 1 to achieve a
better EXIT transfer performance. All these problems deserve further
investigations.



%
\appendices
\section{EXIT function of ESE}\label{app:ese}
Using (\ref{eq:receive}) and (\ref{eq:ESE}), the EXIT function of
the ESE is derived as
\begin{eqnarray}
E[x_t^{(k)}L^e(x_t^{(k)})]\!\!\!\!\!\!\!&&=E[\frac{2\sqrt{\frac{E_b}{L}}x_t^{(k)}\left(y_t-\sqrt{\frac{E_b}{L}}\sum_{i=1,i\neq
k}^K\tanh(\frac{L^{a}(x^{(i)}_t)}{2})\right)}{\frac{E_b}{L}\sum_{i=1,i\neq
k}^K\left(1-(\tanh(\frac{L^{a}(x^{(i)}_t)}{2}))^2\right)+\frac{N_0}{2}}]\nonumber\\
&&=E[\frac{2\frac{E_b}{L}+2\frac{E_b}{L}x_t^{(k)}\sum_{i=1,i\neq
k}^K\left(x_t^{(i)}-\tanh(\frac{L^{a}(x^{(i)}_t)}{2})\right)+2\sqrt{\frac{E_b}{L}}x_t^{(k)}z_t}{\frac{E_b}{L}\sum_{i=1,i\neq
k}^K\left(1-(\tanh(\frac{L^{a}(x^{(i)}_t)}{2}))^2\right)+\frac{N_0}{2}}]\nonumber\\
&&=E[\frac{4}{2\sum_{i=1,i\neq
k}^K\left(1-(\tanh(\frac{x^{(i)}_tL^{a}(x^{(i)}_t)}{2}))^2\right)+{L}/{\frac{E_b}{N_0}}}]\label{eq:ESEXIT1}\\
&&=E[\frac{4}{2\sum_{i=1}^{K-1}\left(1-(\tanh
(\hbar_i))^2\right)+{L}/{\frac{E_b}{N_0}}}]\triangleq
\phi(m_a,\frac{E_b}{N_0})\nonumber
\end{eqnarray}
where $\hbar_i, i=1,...,K-1$ are i.i.d. with $\hbar_i\sim \mathcal
{N}(\frac{m_a}{2},\frac{m_a}{2})$. Eq.~(\ref{eq:ESEXIT1}) is due to
that $x^{(k)}_t\sim \mathcal {U}(\mathcal {X})$ is independent of
$x^{(i)}_t, i\neq k$, and $z_t$. Note that
$\phi(m_a,\frac{E_b}{N_0})$ is an average EXIT function of the ESE
for all the possible transmitted symbols.

We give an upper bound for $\phi(m_a,\frac{E_b}{N_0})$
\begin{lemma} \label{lem:ESElim}
\begin{eqnarray}
\phi(m_a,\frac{E_b}{N_0})\leq4\frac{E_b}{LN_0}\nonumber
\end{eqnarray}
with equality if $m_a\rightarrow+\infty$.\myQED
\end{lemma}
\emph{Proof:} The upper bound is from the fact $|\tanh(x)|\leq 1$.
We prove the equality in the upper bound. Let $\frac{1}{2}<p<1$ be a
constant. We have
\begin{eqnarray}
\lim_{m_a\rightarrow
+\infty}\phi(m_a,\frac{E_b}{N_0})\!\!\!\!\!\!&&\geq
\lim_{m_a\rightarrow
+\infty}\frac{4\prod_{i=1}^{K-1}\textrm{Pr}(\hbar_i\geq
m_a-(m_a)^p)}{2\sum_{i=1}^{K-1}\left(1-(\tanh
(m_a-(m_a)^p))^2\right)+{L}/{\frac{E_b}{N_0}}}\label{eq:lemma2-1}\\
&&\geq \lim_{m_a\rightarrow
+\infty}\frac{4(1-\frac{m_a}{2(m_a)^{2p}})^{K-1}}{2\sum_{i=1}^{K-1}\left(1-(\tanh
(m_a-(m_a)^p))^2\right)+{L}/{\frac{E_b}{N_0}}}\label{eq:lemma2-2}\\
&&=4\frac{E_b}{LN_0}\nonumber
\end{eqnarray}
where (\ref{eq:lemma2-1}) is from monotone increasing of $\tanh(x)$
for $x\geq0$, and (\ref{eq:lemma2-2}) is from Chebyshev inequality.
Using the squeeze theorem we obtain $\lim_{m_a\rightarrow
+\infty}\phi(m_a,\frac{E_b}{N_0})=4\frac{E_b}{LN_0}$. The lemma is
proved.


\ifCLASSOPTIONcaptionsoff
  \newpage
\fi


\begin{thebibliography}{1}

\bibitem{verylowrate} A. J. Viterbi
``Very low rate convolutinal codes for maximum theoretical
performance of spread-spectrum multiple-access channels," \emph{IEEE
J. Sel. Areas Commun.}, vol. 8, no. 4, pp. 641-649, May 1990.

\bibitem{cdma} A. J. Viterbi, \emph{CDMA: Principles of Spread Spectrum Communication}, Reading, MA, Addison-Wesley, 1995.

\bibitem{IDMA} W. Leung, L. Liu, and P. Li,
``Interleaving-based multiple access and iterative chip-by-chip
multi-user detection," \emph{IEICE Trans. Commun.}, vol. E86-B, no.
12, pp. 3634-3637, Dec. 2003.

\bibitem{idmaj} P. Li, L. Liu, K. Wu, and W. Leung, ``Interleaving-division multiple-access,"
\emph{IEEE Trans. Wireless Commun.}, vol. 5, no. 4, pp. 938-947,
Apr. 2006.

\bibitem{analysisIDMA}K. Li, X. Wang, and P. Li, ``Analysis and optimization of interleave-division
multiple-access communication systems," \emph{IEEE Trans. Wireless
Commun.}, vol. 6, no. 5, pp. 1973-1983, May 2007.

\bibitem{spreaddesign} G. Song, J. Cheng, and W. Yoichiro,
``Spreading and interleaving design for synchronous
interleave-division multiple-access," \emph{IEICE Trans.
Fundamentals.}, vol. E95-A, no. 3, pp. 646-656, Mar. 2012.
















\bibitem{gaussian} S. Y. Chung, T. J. Richardson, and R. L. Urbanke, ``Analysis of sum-product decoding of low-density
parity-check codes using a Gaussian approximation," \emph{IEEE
Trans. Inf. Theory}, vol. 47, no. 2, pp. 657-670, Feb. 2001.

\bibitem{Convergence} S. ten Brink, ``Convergence behavior of iteratively decoded
parallel concatenated codes," \emph{IEEE Trans. Commun.}, vol.~49,
no. 10, pp. 1727-1737, Oct. 2001.

\bibitem{modern} T. Richardson and R. Urbanke, \emph{Modern Coding Theory}, Cambridge,
Cambridge University Press, 2008.

\bibitem{lin} W. E. Ryan and S. Lin, \emph{Channel Codes: Classical and Modern}, Cambridge,
Cambridge University Press, 2009.

\bibitem{supermap}T. Wo and P. A. Hoeher, ``Universal coding approach for
superposition mapping," \emph{in Proc. 6th International Symposium
on Turbo Codes and Iterative Information Processing (ISTC 2010)},
pp. 324-328, France, Sept. 2010.

\bibitem{supermapj}P. A. Hoeher and T. Wo, ``Superposition modulation: myths and facts," \emph{IEEE Commun. Mag.},
vol. 49, no. 12, pp.110-116, Dec. 2011.


\end{thebibliography}
\end{document}